\documentclass[12pt,letterpaper]{iopart}
\usepackage{iopams}
\usepackage{graphicx}
\usepackage{amssymb}

\newcommand{\bra}[1]{\langle{#1} |}
\newcommand{\ket}[1]{|{#1}\rangle  }

\newcommand{\ketbra}[2]{\vert {#1} \rangle \langle{#2}\vert}
\providecommand{\openone}{\leavevmode\hbox{\small1\kern-3.8pt\normalsize1}}

\begin{document}

\title{Recovering Entanglement by Local Operations}

\author{A. D'Arrigo$^{a,b,c}$, R. Lo Franco$^{c,d}$, G. Benenti$^{e,f}$, 
        E. Paladino$^{b,a,c,g}$ and G. Falci$^{b,a,c,g}$}
\address{$^a$CNR-IMM UOS Universit\`a (MATIS), Consiglio Nazionale delle Ricerche,
         Via Santa Sofia 64, 95123 Catania, Italy}
\address{$^b$Dipartimento di Fisica e Astronomia, Universit\`a degli Studi Catania, 
         Via Santa Sofia 64, 95123 Catania, Italy}
\address{$^c$Centro Siciliano di Fisica Nucleare e Struttura della Materia (CSFNSM),
            Via Santa Sofia 64, 95123 Catania, Italy}
\address{$^d$Dipartimento di Fisica e Chimica, Universit\`a di Palermo, 
             via Archirafi 36, 90123 Palermo, Italy}
\address{$^e$CNISM and Center for Nonlinear and Complex Systems,
            Universit\`a degli Studi dell'Insubria, Via Valleggio 11, 22100 Como, Italy}
\address{$^f$Istituto Nazionale di Fisica Nucleare, Sezione di Milano,
             via Celoria 16, 20133 Milano, Italy}
\address{$^g$Istituto Nazionale di Fisica Nucleare, Sezione di Catania,
            Via Santa Sofia 64, 95123 Catania, Italy}

\begin{abstract}
We investigate the phenomenon of bipartite entanglement
revivals under purely local operations in 
systems subject to local and independent
classical noise sources.  
We explain this apparent paradox in the physical 
ensemble description of the system state by introducing the concept of ``hidden" entanglement, 
which indicates the amount of entanglement that cannot be exploited
due to the lack of classical information on the system.
For this reason this part of entanglement can be recovered without the action of
\textit{non-local} operations or \textit{back-transfer} process.
For two noninteracting qubits under a low-frequency stochastic noise, we show that entanglement 
can be recovered by local pulses only. We also discuss how hidden entanglement may provide new insights about 
entanglement revivals in non-Markovian dynamics.
\end{abstract}

\pacs{03.67.-a, 03.65.Ud, 03.65.Yz}

\maketitle

\section{Introduction}
Entanglement, arguably the most peculiar feature of quantum mechanics,
plays a key role in several quantum information and communication 
applications, including teleportation, quantum dense coding, 
private key distribution, and reduction of communication 
complexity~\cite{nielsen-chuang, benenti-casati-strini, PlenioReview, 
HorodeckiReview}.
To work properly,
all the above tasks generally require pure maximally entangled states.
Since entanglement cannot be generated by Local Operations and Classical 
Communication (LOCC), entangled states 
must be generated somewhere, and then they have to be distributed among different 
parties, possibly far away from each other (\textit{transmission}) \cite{PlenioReview, 
HorodeckiReview}. 
Once entanglement has been distributed, it can be used immediately or 
stored for later use (\textit{storage}).
Systems physically supporting entangled states, unavoidably 
interact with the environment, both during transmission and storage, 
and therefore undergo noisy processes that deteriorate entanglement. 
Quantification of entanglement losses is thereby necessary for all pratical 
purposes.

For a pure state $\rho=|\psi\rangle\langle\psi|$, bipartite entanglement between 
subsystems $A$ and $B$ is unambiguously defined as the \emph{entropy of 
entanglement} $E(|\psi\rangle\langle\psi|)=S(\rho_A)=S(\rho_B)$, where 
$S(\rho_i)$ is the von Neumann entropy of one of the two reduced
states, $\rho_A={\rm Tr}_B \rho$ and $\rho_B={\rm Tr}_A \rho$.
The quantification of entanglement for mixed states is a much more
complicated and still open problem~\cite{PlenioReview,HorodeckiReview}. 
The difficulty roots in the fact that a mixed state $\rho$
may be decomposed into an ensemble of pure states $\rho=\sum_i p_i 
|\psi_i\rangle\langle\psi_i|$, with $p_i>0$ and $\sum_i p_i =1$, in infinite 
different ways.  
The arbitrariness of the decomposition renders any quantification
of mixed-state entanglement cumbersome, since it requires an
optimization over all possible decompositions. 

\begin{figure}[t!]
\begin{center}
\includegraphics[angle=0.0, width=12cm]{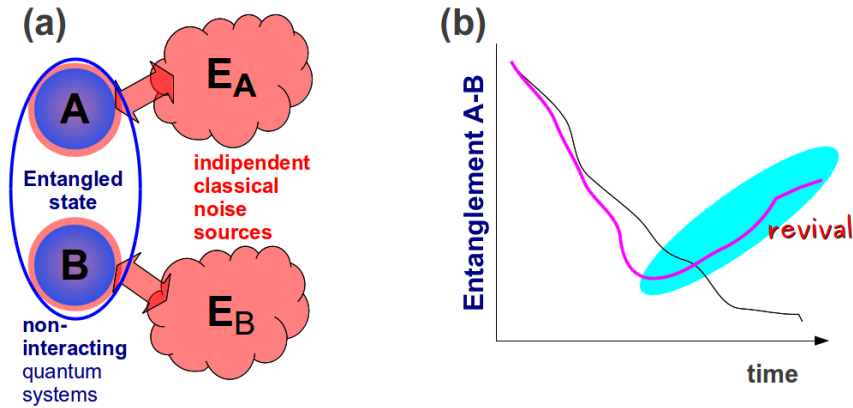}
\end{center}
\caption{(Color online)
(a) Two quantum systems, A and B, initially prepared in an entangled state are transferred to
different locations where they do not interact each other and are subject to
local independent classical noise sources and local operations.
(b) The black (thin) line describes the most usual entanglement behaviour of the considered system.
In this article we will point out the possibility that this system may exhibit a 
non monotonic entanglement behaviour without the action of any non-local control, 
qualitatively sketched by the magenta (thick) line. 
}
\label{fig:simple-sketch}
\end{figure}

In this article, we address the issue of the occurrence of
entanglement revivals of a bipartite system, initially prepared in an entangled state,
when the two subsystems are noninteracting and affected by local independent
classical noise sources and local operations (see Fig.~\ref{fig:simple-sketch}(a)). 
In the absence of non-local operations, entanglement cannot be generated
neither back-transferred to the system from the classical environment.
Nevertheless, during the system dynamics, entanglement quantified by some
measure $E$ may start to increase at some time $\bar{t}$~\cite{lofrancorandom,revivalstochastic}
as illustrated in Fig.~\ref{fig:simple-sketch}(b). 
As we will explain,
the increase of entanglement must be attributed to the manifestation 
of preexisting quantum correlations, that were already present 
before $\bar{t}$. 
The density operator formalism does not capture the presence
of these quantum correlations, thus they are in some sense \textit{hidden}.
Here we point out that the existence of these correlations is enlightened 
if the system is described as a physical ensemble of states
and we introduce the concept of {\em hidden entanglement}.

This paper is structured as follows. In Section~\ref{sec:HE} we introduce a definition 
of hidden entanglement (HE) and 
illustrate the usefulness of this concept by a simple example. 
In Section~\ref{sec:Ent-recovery-Markovian-noise} 
we show that HE between two noninteracting qubits subject to a 
non-Markovian stochastic process can be 
recovered by {\em local} pulses (acting only on one qubit). 
The nature of the observed entanglement revivals and
the relation of this phenomenon with the environment being classical or quantum, 
is clarified.
In Section~\ref{sec:discussion} we critically discuss some key points related to the definition of
HE. In particular, we show that
entanglement recovery does not violate the monotonicity axiom:
\textit{entanglement cannot increases under LOCC}~\cite{PlenioReview,HorodeckiReview,Bennett96}. 
We draw our conclusions in Section~\ref{sec:conclusions}.

\section{Hidden entanglement}
\label{sec:HE}
Let us consider a bipartite system described by an ensemble of states
${\cal A}=\{(p_i,|\psi_i\rangle)\}$.  That is, 
we know the statistical distribution of the bipartite pure states
$\{|\psi_i\rangle\}$, occurring with probabilities
$\{p_i\}$, so that $\rho=\sum_i p_i |\psi_i\rangle\langle \psi_i|$,
but the state of any individual system in the
ensemble is unknown. 
The average entanglement of ${\cal A}$ is defined as
~\cite{Bennett96,Cohen98,Nhal04,Carvalho07}:
\begin{equation} 
   {E}_{av}({\cal A})=\sum_i p_i E(|\psi_i\rangle\langle \psi_i|).
\label{eq:averageEntanglement}
\end{equation}
If each system in the ensemble evolves during time $t$ under
LOCC, the maximum amount of entanglement of the corresponding density 
operator $\rho(t)$ can never overcome the initial value ${E}_{av}({\cal A})$.   
This statement can be proved by the following simple argument.
Suppose Charlie prepares a bipartite system in a (possibly entangled) pure state
of the ensemble  ${\cal A}$.
Then he sends one half of the system to Alice and 
the other half to Bob
through noiseless quantum channels.  
Alice and Bob communicate only by a noiseless classical channel. 
Charlie repeats this operation $N$ times. 
Among these, a certain number $M_j$ of times Alice and Bob deal 
with the state $\ket{\psi_{j}}$.
If Alice (or Bob) receives from Charlie the classical information about 
which state he sent each time, in the limit of large $N$
Alice and Bob can distil - by only using LOCC - up to 
$M_{j}\, E(\ket{\psi_{j}} \bra{\psi_{j}})$ 
maximally entangled states from the  
$M_j$ states $\ket{\psi_{j}}$ at their disposal~\cite{Bennett96}. 
Distillable entanglement is in fact the entropy 
of entanglement for pure bipartite states.
Therefore, the maximum entanglement that Alice and Bob 
can distill per pair, by using classical information from Charlie, is
\begin{equation}  
 \lim_{N \to \infty} \frac{1}{N}\sum_i M_i \, 
E(\ket{\psi_i}\bra{\psi_i})= \sum_i p_i \, 
E(\ket{\psi_i}\bra{\psi_i}),
\label{eq:MaximumDistillableEntanglement}
\end{equation}
which is just the average entanglement of Eq.~(\ref{eq:averageEntanglement})
($\lim_{N \to \infty} \frac{M_i}{N} \equiv p_i$). 
\begin{figure}[t!]
\begin{center}
\includegraphics[angle=0.0, width=12.5cm]{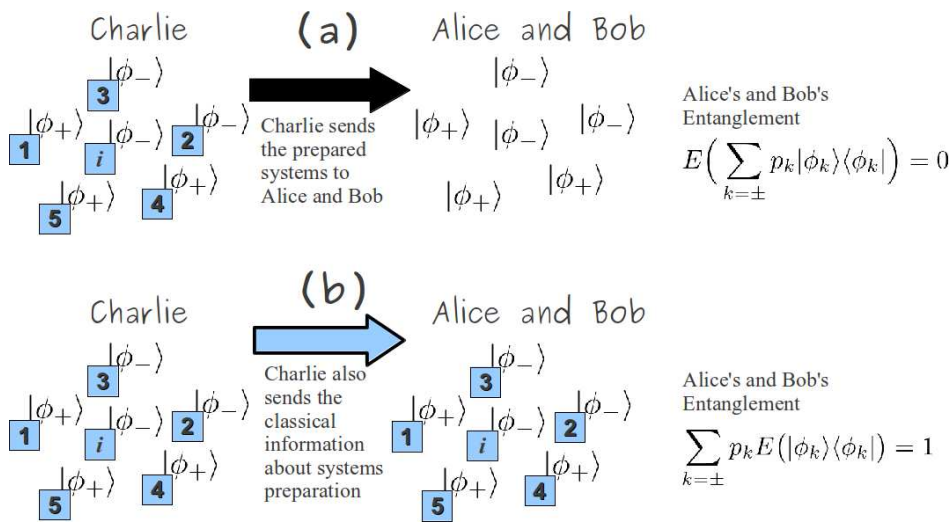}
\end{center}
\caption{(Color online)
Pictorial illustration of the concept of hidden entanglement.
Charlie prepares a large number of bipartite 
systems in the pure states 
$\ket{\psi_i}$, as described by the quantum
ensemble ${\cal A}=\{(p_i,|\psi_i\rangle)\}$. Here, for the sake of simplicity, we assume that
$\ket{\psi_i}$ can be chosen as $\ket{\phi_\pm}=(\ket{00}\pm\ket{11})/\sqrt{2}$
 with the same probability~\cite{Eisert00}.
\textbf{(a)} Charlie sends one half of each system to Alice and the other half to Bob through 
noiseless quantum channels.
The entanglement Alice and Bob can distill per pair vanishes, $E(\rho_{AB})$=0,
since Alice's and Bob's state 
$\rho_{AB}=\sum_i p_i \ketbra{\psi_i}{\psi_i}=\frac{1}{2}\,(\ket{00}\bra{00}+
\ket{11}\bra{11})$ is separable.
\textbf{(b)} Charlie uses a classical telephone line to communicate
the states preparation to Alice. The entanglement Alice and Bob can now
distill per pair is equal to 1
(Alice can perform a phase flip on her qubit, each time she 
knows that the corresponding pair is $\ket{\phi_-}$, so that all Alice's and Bob's pairs 
at the end are in the state $\ket{\phi_+}$).
In the two scenarios, Alice and Bob \textit{physically share the same system}.
Here the root of entanglement recovery lies in the acquisition of {\em classical} information.
Since this occurs in the absence of any interaction between the quantum systems or entanglement transfer 
through a third quantum system, the phenomenon is entirely due to the manifestation of quantum
correlations already present in the system and in this sense ``hidden''.
}
\label{fig:CharliePicture}
\end{figure}

We define the {\it hidden entanglement} (HE)
of the ensemble ${\cal A}=\{(p_i,|\psi_i\rangle)\}$ as
the difference between the average entanglement of the ensemble and the 
entanglement~\cite{PlenioReview,HorodeckiReview} 
of the state $\rho=\sum_i p_i |\psi_i\rangle\langle\psi_i|$
quantified by any convex measurement $E(\rho)$
(reducing to the entropy of entanglement for pure states), that 
is\footnote{Note that in Ref.~\cite{Cohen98} the expression
``hidden entanglement'' is used with a different meaning.}
\begin{equation}
\begin{array}{l}
{\displaystyle
    {E_h}({\cal A}) \equiv E_{av}({\cal A}) - E(\rho)\,=
}
\\
{\displaystyle\hspace{0.5cm}
= \sum_i p_i E(|\psi_i\rangle\langle\psi_i|)-
E\left(\sum_i p_i |\psi_i\rangle\langle\psi_i|\right).
}
    \label{eq:MeasHiddenEntang}
\end{array}
\end{equation} 
Due to convexity, $E_h$ is always larger than or equal to zero.
The meaning of HE Eq.~(\ref{eq:MeasHiddenEntang}) is clear: 
It is the entanglement 
that cannot be exploited as a resource due to the
\emph{lack of knowledge} about which state of the mixture 
we are dealing with (see Fig.~\ref{fig:CharliePicture}). Such entanglement can be recovered
(unlocked~\cite{Cohen98,Gour07,Eisert00}) once this classical information is
provided, \textit{without the help of any non local operation}.
We remark, as it is clear from the definition Eq.~(\ref{eq:MeasHiddenEntang}), that HE is associated to 
the specific quantum ensemble description of the system state. 
We will refer to situations where the system dynamics 
\textit{admits a single physical decomposition in terms 
of an ensemble of pure state evolutions}. This is always possible, at least in principle, when the 
system is affected by \textit{classical noise sources}, as illustrated in 
\ref{quantum-ensemble-time-evolutions}. 

In the rest of this article we will illustrate the meaning of HE,
expressed by Eq.~(\ref{eq:MeasHiddenEntang}), with various examples.
There exist several inequivalent measures of mixed state entanglement~\cite{PlenioReview,HorodeckiReview}. 
Here we consider the entanglement of formation $E_f(\rho)$, which  is an upper bound for any 
bipartite entanglement measure~\cite{horodeckiprl}, so that $E_{av}-E_f(\rho)$ is a lower bound for the hidden 
entanglement. $E_f(\rho)$ can be readily computed for two-qubit systems via 
the concurrence $C(\rho)$~\cite{Wootters98}.

\subsection{Entanglement revivals under random local fields}
We first illustrate the concept of HE by considering
a random, local dynamics and demonstrating that, under proper conditions,
a {\em complete recovery} of the entanglement $E_f(\rho)$ may occur. 

A basic property of the average entanglement is its invariance
under local unitary transformations.
In particular this is the case of the evolution 
in a random local external field \cite{alickibook}
inducing \emph{local random unitaries} $U_\alpha(t)\otimes V_\beta(t)$ on a bipartite system,
with the operators $U_\alpha$ and $V_\beta$ acting respectively on the first
and on the second subsystem, and depending on the random 
variables $\alpha,\beta$. 
Let us suppose that the system is initially prepared in a pure state $\ket{\varphi(t_0)}=\ket{\varphi_0}$.
Thus at any subsequent time the system is
described by the quantum ensemble 
${\cal A}(t)=\{(p_{\alpha\beta},|\varphi_{\alpha\beta}(t)\rangle)\}$, 
with $|\varphi_{\alpha\beta}(t)\rangle = (U_\alpha(t)\otimes V_\beta(t)) \ket{\varphi_0}$.
The average entanglement of the ensemble $\cal A$ is conserved by 
this dynamics, $E_{av}({\cal A}(t))=E_{av}({\cal A}(t_0))$. 
On the other hand, the entanglement of the mixture 
$\rho(t)=\sum_{\alpha,\beta}p_{\alpha\beta} 
\ketbra{\varphi_{\alpha\beta}(t)}{\varphi_{\alpha\beta}(t)}$ is only upper bounded 
by the average entanglement: $E_f(\rho(t))\leq E_{av}({\cal A}(t))$,
implying that a variable (time dependent) HE may exist.
This is clearly illustrated by the following simple example.

Let us consider a two-qubit system $AB$ initially
prepared in the maximally entangled Bell state $|\phi^+\rangle$.
The time evolution consists of local unitaries, but we have no 
complete information about which local unitary is acting. In particular, we suppose that
the qubit $A$ undergoes, with equal probability, 
a rotation about the $x$-axis of its Bloch sphere, 
$U_x(t)=\mathrm{e}^{-\mathrm{i} \sigma_x \omega t/2}$,
or a rotation around the $z$-axis, $U_z(t)=\mathrm{e}^{-\mathrm{i} \sigma_z \omega t/2}$,
while the qubit $B$ remains unchanged.
Hence, the ensemble ${\cal A}$ at time $t$ is
\begin{equation}
{\cal A}(t)=\left\{\left(\frac{1}{2}, (U_x(t)\otimes \openone_B) 
\ket{\phi^+}\right),
\,\left(\frac{1}{2}, (U_z(t) \otimes \openone_B) \ket{\phi^+}\right)\right\}.
\end{equation}
Since we are dealing with random local unitaries, the average 
entanglement of $\cal A$ is constant in time, $E_{av}({\cal A}(t))=1$.
On the other hand, the entanglement of the state $\rho(t)$ changes in time. 
At $\overline{t}=\frac{\pi}{\omega}$,
$\rho(\overline{t})=\frac{1}{2}|\phi^-\rangle\langle\phi^-|+
\frac{1}{2}|\psi^+\rangle\langle\psi^+|$ is separable, whereas
at $2 \overline{t}$, $U_x(2 \overline{t})=U_z(2 \overline{t})=\openone_A$ and 
the initial maximally entangled state is recovered
\footnote{We use the notation $|\psi^\pm\rangle= (\ket{01}\pm\ket{10})/\sqrt{2}$}.
In the interval $[\overline{t},2\overline{t}]$ the entanglement revives from
zero to one without the action of any nonlocal quantum operation, 
thus apparently violating the monotonicity axiom.
The ensemble description tells us that at time
$\overline{t}$ the system is always in an entangled state 
($\ket{\phi_-}$ or $\ket{\psi_+}$), but the lack of knowledge about 
which local operation the system underwent prevents us from distilling 
any entanglement: entanglement is {\em hidden}, 
${E_h}({\cal A}(\overline{t}))=1$ and $E_f(\rho(\overline{t}))=0$.
At time $2\overline{t}$ this lack of knowledge
is irrelevant since the two possible time evolutions result
in the identity operation $\openone_A$ and entanglement is recovered, ${E_h}({\cal A}(2\overline{t}))=0$ 
and $E_f(\rho(2\overline{t}))=1$.

We notice that entanglement revivals 
under random local fields have been studied in Ref.~\cite{lofrancorandom}.
Here we explain and quantify this phenomenon in terms of HE. 

\section{Entanglement recovery in the presence of classical non-Markovian noise}
\label{sec:Ent-recovery-Markovian-noise}
A fingerprint of the existence of HE is the possibility to completely recover entanglement
of a noisy bipartite system by the action of {\em local} pulses. Here we consider a simple
system consisting of two noninteracting qubits affected by classical  
non-Markovian noise.
This simplified model captures essential features of several nanodevices
whose dynamics is dominated by low-frequency 
noise~\cite{FalciPrl05,Ithier2005,Bylander,Chiarello}. 
We suppose the two qubits are initially 
prepared in a Bell state $\ket{\varphi_0}$
and, for the sake of simplicity, assume that only qubit $A$ is affected 
by phase noise (pure dephasing), as described by ($\hbar=1$)  
\begin{equation}\label{hamiltonian-Echo}
{\cal H}_A(t)=  [-\Omega_A \sigma_{z} +
\varepsilon(t) \sigma_{z} + \mathcal{V}(t) \sigma_{x}]/2 \, ,
\end{equation}
where $\varepsilon(t)$ is a stochastic process, and $\mathcal{V}(t)$ an 
external control field.
Qubit $B$ evolves unitarily under a Hamiltonian ${\cal H}_B(t)$.

\begin{figure}[t!]
\includegraphics[angle=0.0, width=6.7cm]{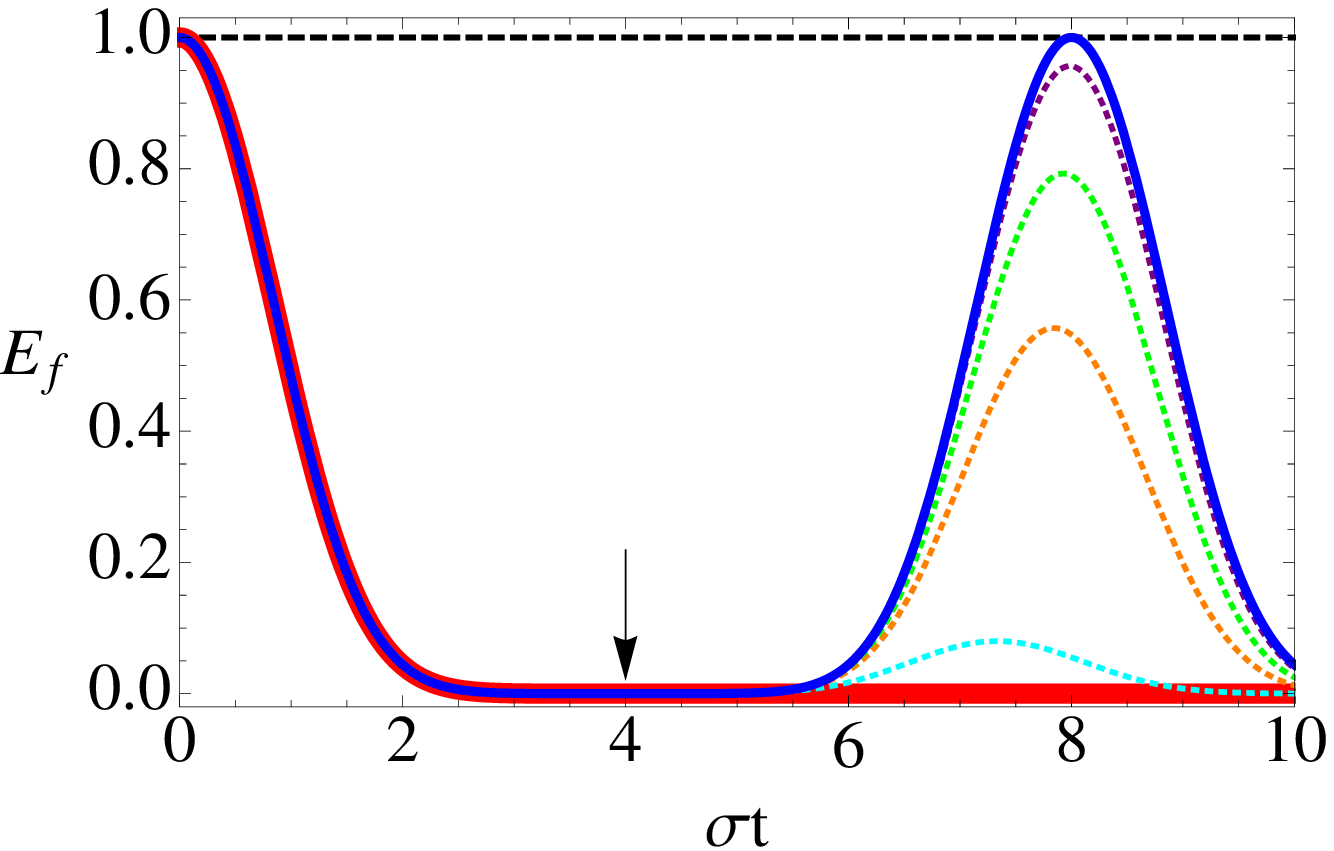}
\includegraphics[angle=0.0, width=6.7cm]{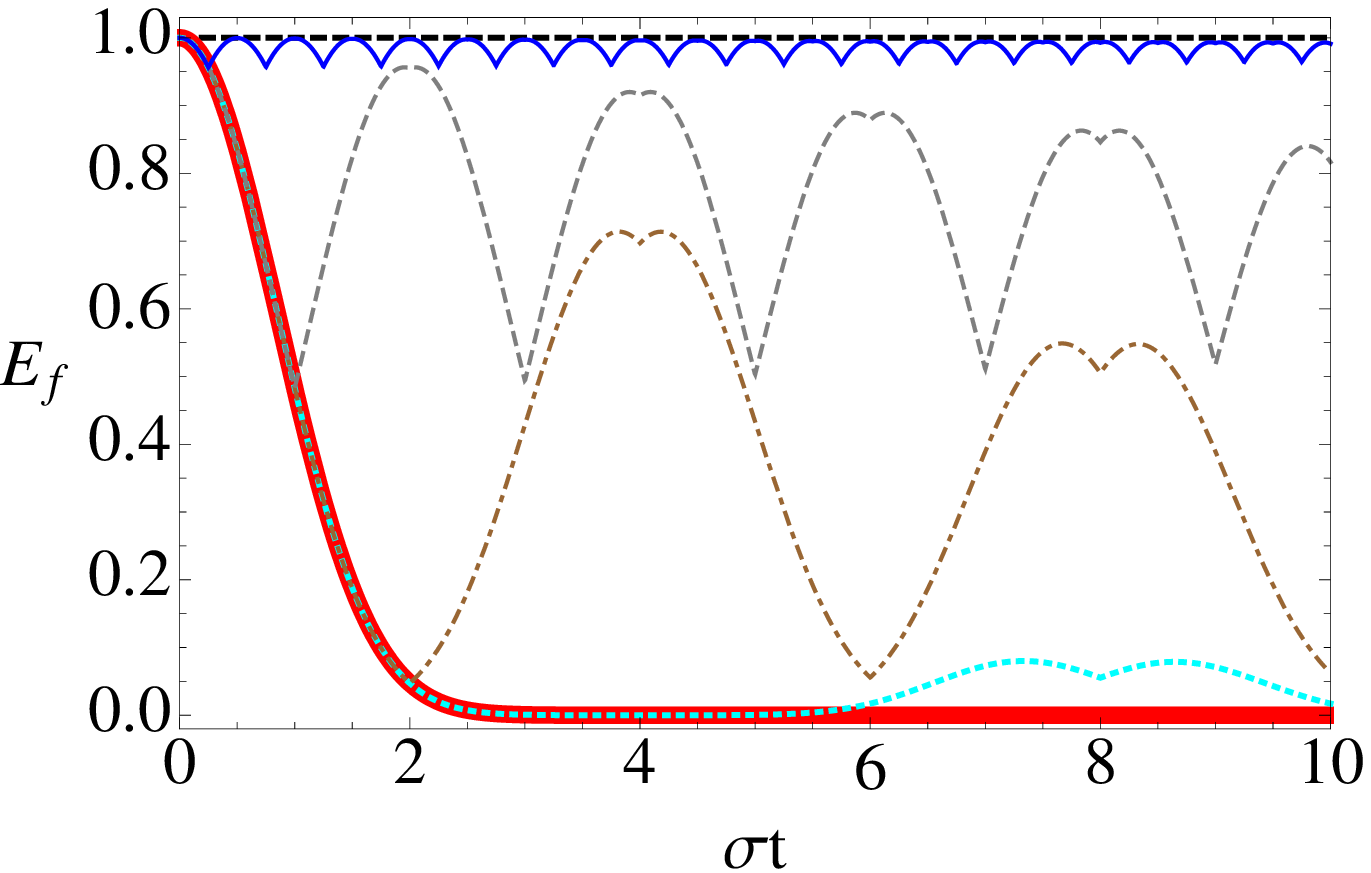}
\caption{(Color online)
Entanglement of formation $E_f(\rho(t))$ as a function
of the dimensionless time $\sigma t$. 
Left panel: The thick (red) curve corresponds to the free evolution in the presence of
static noise, the thin (blue) solid curve is the result of the echo pulse applied at time $\sigma\overline{t}=4$ (indicated 
by the arrow), Eq.~(\ref{twoqubitconc}).
The dashed line is the system average  entanglement $E_{av}({\cal A}(t))=1$.
Dotted curves represent $E_f(\rho(t))$
for a $\varepsilon(t)$ with a Lorentzian power spectrum  when an echo pulse is applied at time $\sigma \bar t =4$,
from Eqs.~(\ref{eq:Concurrence-generic-Gaussian-noise}) and (\ref{eq:FilterFunction-2}).
From bottom to top: $\sigma\tau=20$ (cyan curve), $100$ (orange curve), $200$ (green curve),
and $500$ (purple curve).
Perfect recovery is obtained in  the limit $\tau/\overline{t}\to\infty$, corresponding 
to static noise (blue thin solid curve). 
Right panel: 
The (red) thick solid curve corresponds to $E_f(\rho(t))$ 
evaluated for a stochastic process $\varepsilon(t)$
with a Lorentzian power spectrum and correlation time $\sigma\tau=20$,
in the case of free evolution, from Eqs.~(\ref{eq:Concurrence-generic-Gaussian-noise}) 
and (\ref{eq:FilterFunction-1}).
The other curves refer to a PDD protocol  applied to qubit A
with equally spaced $\pi$-pulses, applied at times
$t_k=k\Delta t$. 
$E_f(\rho(t))$ is numerically evaluated from 
Eqs.~(\ref{eq:Concurrence-generic-Gaussian-noise}) and (\ref{eq:FilterFunction-3}):
$\tau/\Delta t=5$ for the dotted (cyan) curve,
$\tau/\Delta t=10$ for the dot-dashed (brown) curve,
$\tau/\Delta t=20$ for the dashed (gray) curve and
$\tau/\Delta t=80$ for the thin solid (blue) curve.
Almost perfect recovery is obtained when $\tau/\Delta t\gg 1$.
}
\label{fig:PureDeph-echo}
\end{figure}

To start with, we suppose that $\varepsilon(t)$ is sufficiently slow
to be considered {\em static} during the evolution time $t$, with a value randomly fluctuating from one quantum evolution 
to the other. We assume that  
$\varepsilon$ is a Gaussian random variable 
with zero expectation value and standard deviation $\sigma$. $\mathcal{V}(t)$ indicates a hard
echo $\pi$-pulse at time $\overline{t}$, short enough 
to neglect the effect of noise during its application. The evolution operator during the pulse 
is $\mathrm{e}^{-\mathrm{i}\sigma_x\pi/2}=-\mathrm{i}\sigma_{x}$. 
Static noise~\cite{FalciPrl05,Ithier2005,Bylander,Chiarello} 
produces an effect analogous 
to inhomogeneous broadening in nuclear magnetic 
resonance (NMR)~\cite{Slichter}.
The system dynamics is described by the quantum ensemble
${\cal A}(t)=\{p(\varepsilon)d\varepsilon,\,\ket{\varphi_\varepsilon(t)}\}$,
where $\ket{\varphi_\varepsilon(t)}=
\hat{T}\mathrm{e}^{-\mathrm{i}\int_0^t{{\cal H}_A(t')dt'}}\otimes \hat{T}\mathrm{e}^{-\mathrm{i}\int_0^t{{\cal H}_B(t')dt'}}   
\ket{\varphi_0}$ 
and $p(\varepsilon)$ is the Gaussian probability density function of $\varepsilon$
(see \ref{quantum-ensemble-time-evolutions}).
Note that for each realization of the stochastic process $\varepsilon(t)$
the system state acquires a random phase.
In the density operator description of the system, the information
about such random phase is 
lost by averaging the evolved pure state $\ket{\varphi_\varepsilon(t)}$ 
with respect to the random variable $\varepsilon$: 
$\rho(t)=\int d\varepsilon p(\varepsilon) |\varphi_\varepsilon(t)\rangle\langle \varphi_\varepsilon(t)|$.
A system prepared
in a Bell state $\ket{\varphi_0}$ evolves in a mixture whose 
concurrence $C(\rho(t))$ is twice the absolute value of the only 
non-zero coherences, and reads 
\begin{equation}
C(\rho(t))=
\left\{
\begin{array}{ll}
\mathrm{e}^{-\frac{1}{2}\sigma^2 t^2},& 0\leq t\leq\overline{t},\\
\mathrm{e}^{-\frac{1}{2}\sigma^2 (t-2\overline{t})^2},& \overline{t}\leq t\leq 2\overline{t} \, .
\end{array}
\right .
\label{twoqubitconc}
\end{equation}
The entanglement of formation $E_f(\rho(t))$
is obtained directly from $C(\rho(t))$~\cite{Wootters98}:
\begin{equation}
E_f(\rho(t))=
\mathtt{h}\Big(\frac{1+\sqrt{1-C(\rho(t))^2}}{2}\Big), 
\label{eq:wootters-formula}
\end{equation}
where $\mathtt{h}(x)=-x\log_2x-(1-x)\log_2(1-x)$.
In absence of pulses, $E_f$  decays and almost vanishes, $E_f(\rho(t))\simeq 0$, at times $\sigma t\gg 1$,
see Fig.~\ref{fig:PureDeph-echo}, left panel, thick (red) curve. 
Differently, after the action of a local pulse at $t=\overline{t}$
the entanglement increases 
reaching at $t=2\overline{t}$ its initial maximum value 
$E_f(\rho(2\overline{t}))=E_f^\textrm{max}= 1$ (thin blue curve).
This is exactly the average entanglement (dashed line)
of the evolved physical ensemble ${\cal A}$, 
$E_{av}({\cal A}(t))=1$. Indeed for each realization of 
$\varepsilon(t)$, the system 
remains in a pure maximally entangled state.

In the general case of a Gaussian stochastic process $\varepsilon(t)$
the concurrence can be expressed in the form
\cite{Bylander,FilterFunctions}
\begin{equation}
   C(\rho(t))=e^{-\frac{1}{2}\int_{-\infty}^{+\infty}\frac{d\omega}{2\pi}\,S(\omega) \frac{F(\omega,t)}{\omega^2}}
\label{eq:Concurrence-generic-Gaussian-noise}
\end{equation}
where $S(\omega)=\int_{-\infty}^{+\infty}dt e^{-i \omega t }\langle \varepsilon (t) \varepsilon (0) \rangle$
is the power spectrum of the process $\varepsilon(t)$. The function $F(\omega,t)$
represents a filter function~\cite{Bylander,FilterFunctions} depending on the system unitary evolution.
When the system freely evolves under ${\cal H}_A(t)$ in the absence of external control
actions ($\mathcal{V}(t)=0$ in (\ref{hamiltonian-Echo})), the filter function
$F_{free}(\omega,t)$ reads
\begin{equation}
F_{free}(\omega,t)=4\sin^2 \Big (\frac{\omega t}{2} \Big )   \, .
\label{eq:FilterFunction-1}
\end{equation} 
For an echo protocol, with a $\pi$-pulse at time $\overline{t}$ the
filter function reads \cite{Ithier2005}
\begin{equation}
 F_{echo}(\omega,t) = 4\Big[\sin^2 \frac{\omega \bar{t}}{2}
                +\sin^2\frac{\omega(t-\bar{t})}{2} 
                 -2 \cos \frac{\omega t}{2} \sin\frac{\omega\bar{t}}{2} 
                    \sin\frac{\omega(t-\bar{t})}{2} \Big] \,. 
\label{eq:FilterFunction-2}
\end{equation}
We consider a stochastic process with an exponential autocorrelation function,
$\langle \varepsilon (t) \varepsilon (0) \rangle=\sigma^2 \mathrm{e}^{-|t|/\tau}$,
with noise correlation time $\tau$.
Also in this case, a local echo pulse leads to a significant entanglement recovery 
provided that the noise correlation time is sufficiently large,
$\tau \gg \overline{t}$
(Fig.~\ref{fig:PureDeph-echo}, left panel, dotted lines). 
Better performances can be achieved applying a train of pulses.
For a periodic dynamical decoupling  (PDD) protocol, i. e. a sequence of 
$\pi$-pulses applied at equally spaced times $t_k= k \Delta t$, 
the concurrence takes the form  (\ref{eq:Concurrence-generic-Gaussian-noise}) with the filter function \cite{Uhrig}
\begin{eqnarray}
 &&\hspace{-1cm} F_{PDD}^{(\Delta t)}(\omega,t)=\Big|1+
    (-1)^{\bar{n}}e^{i\omega t}+2\sum_{k=1}^{\bar{n}}(-1)^ke^{i\omega k \Delta t}\Big|^2,
\label{eq:FilterFunction-3}
\end{eqnarray}
where $\bar{n}$ denotes the integer part of $\frac{t}{\Delta t}$. 
In this case, the recovery improves with increasing 
the ratio $\tau/\Delta t$ between the noise correlation time and the time interval between consecutive pulses,
see Fig.~\ref{fig:PureDeph-echo}, right panel.

These examples show the possibility to \emph{fully recover  the entanglement $E(\rho)$ by a local operation}.
The physical mechanism behind this phenomenon is very simple: a local 
$\pi$-pulse applied at some time $\bar{t}$ refocuses the different qubit 
quantum evolutions restoring at time $2 \overline t$ the qubit $A$ coherence and
consequently (qubit $B$ evolves unitarily) causing the entanglement to reappear, 
with an efficiency depending on the correlation time of the stochastic process.
Note that the non-Markovian nature of the stochastic process
is a necessary (but not sufficient) condition for the non-monotonous entanglement behaviour.
Indeed, to observe revivals the environment must keep memory of its states on a time scale
larger than the system evolution time. Under this condition, after the pulse there is a ``back-flow"~\cite{breuer}  
of the {\em classical information}
on the system's {\em phase} which the environment has acquired during the evolution before the pulse. 

Entanglement revival after the application of the pulses may appear paradoxical
at first sight. Entanglement is by definition a nonlocal resource, whereas
we only acted locally on one qubit, without any transfer of entanglement
from the environment which is a classical noise source. 
The key point is that here entanglement is not destroyed during the time evolution,
as indicated by the average entanglement of the ensemble ${\cal A}$ describing
the system dynamics, which is maximum at any time $E_{av}({\cal A}(t))=1$.
Entanglement is instead {\em hidden}: because of the lack of classical knowledge on the system state
due to defocusing among the different evolutions of the (maximally entangled) states of ${\cal A}(t)$,
the HE grows before the pulse is applied. After the pulse, this 
lack classical knowledge is gradually reduced, vanishing in the limit 
$\tau/\bar{t}\to \infty$ (echo) or $\tau/\Delta t\to \infty$ (PDD).
This is the reason why entanglement can be recovered without any nonlocal control.

\subsection{Nature of entanglement revivals} 
The phenomenon  of entanglement revivals we have examined is conceptually 
different from the revivals that a system can exhibit due to the interaction 
with a non-Markovian {\em quantum} environment.  
Indeed, in this last case, system and environment can also develop quantum correlations, and
entanglement revivals may originate from a different physical mechanism.
To exemplify the conceptual difference between these two situations 
here we consider a fully quantum system 
where the entanglement dynamics cyclically decreases, vanishes at a time 
$\bar{t}$ and then increases, analogously to the case of two qubits in  random local fields.

\begin{figure}[t!]
\includegraphics[angle=0.0, width=7.2cm]{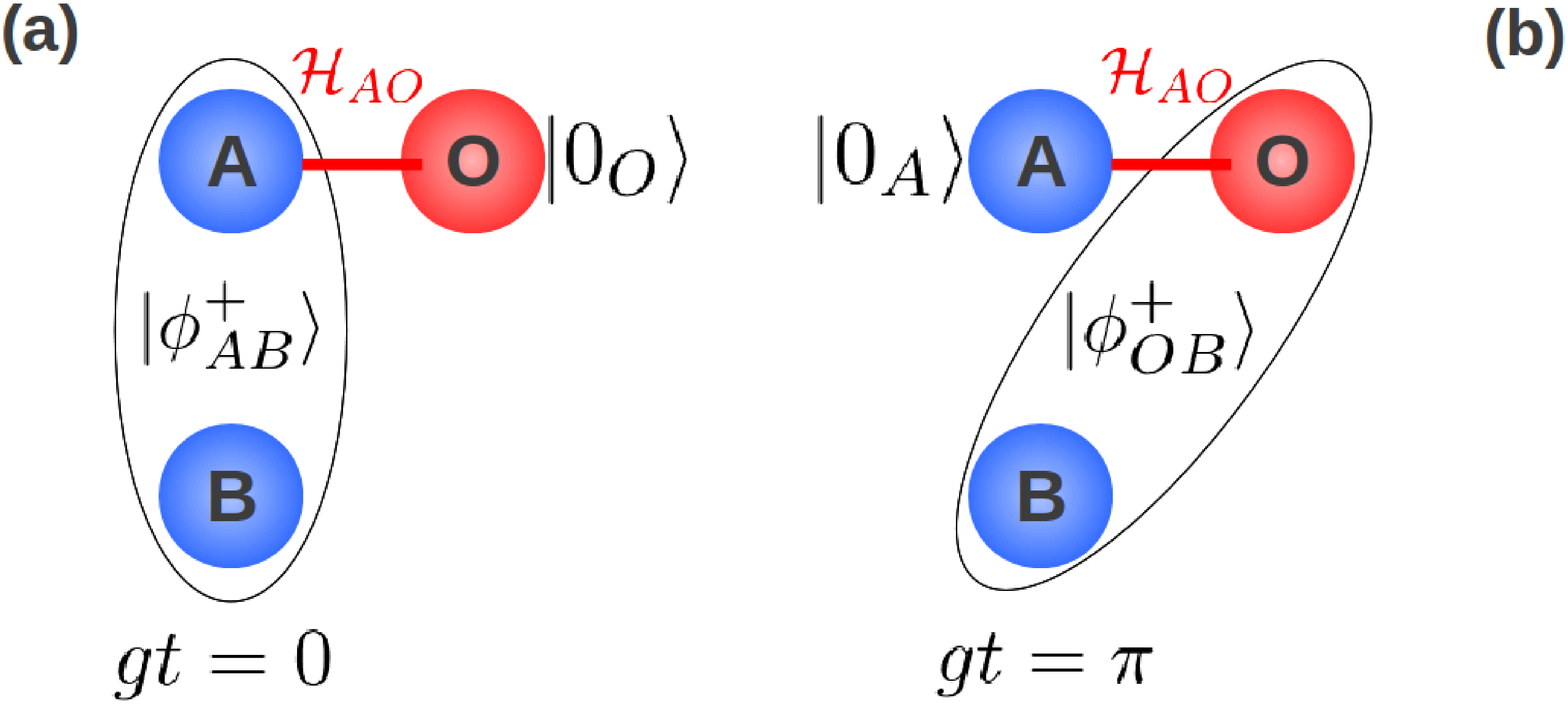}
\includegraphics[angle=0.0, width=6.2cm]{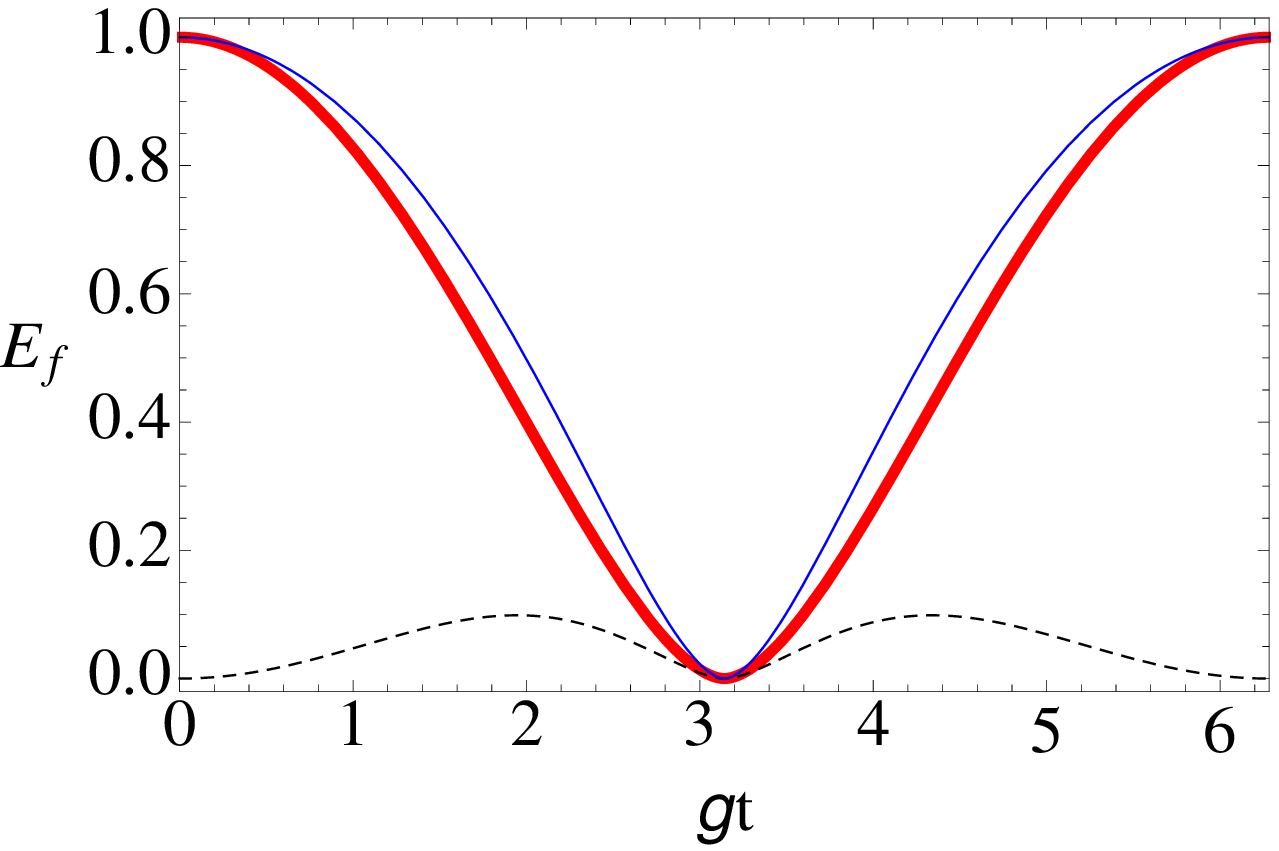}
\caption{(Color online)
Panel \textbf{(a)}: Two qubits are initially ($gt=0$)
prepared in the Bell state $\ket{\phi_{AB}^+}=(\ket{00}+\ket{11})/\sqrt{2}$. Qubit
$B$ is virtually isolated from any environment. Qubit $A$ resonantly interacts
with an harmonic oscillator $O$ via a Jaynes Cummings Hamiltonian. The oscillator
is initially in its ground state $\ket{0_O}$. 
At time $gt=\pi$, because of the interaction between $A$ and $O$, the states of the two systems are swapped.
Panel \textbf{(b)}: Entanglement of formation $E_f(\rho_{AB}(t))$ (thick red line),
average entanglement  $E_{av}({\cal A}(t))$ from Eq.~(\ref{eq:JC-average-entanglement}) relative to 
$AB$ quantum ensemble Eqs.~(\ref{eq:JC-QE-probabilities})-(\ref{eq:JC-QE-states}) (thin blue line)
and corresponding hidden entanglement Eq.~(\ref{eq:JC-hidden-entanglement}) (dashed line) 
as a function of the dimensionless time $g t$.
The vanishing $AB$ entanglement at $gt=\pi$ is due to entanglement transfer to the
$BO$ system,  not to the lack of any classical information on the system $AB$. 
}
\label{fig:JC-interaction}
\end{figure}

Let us consider a two-qubit system $A$-$B$ where $A$ resonantly interacts with a quantum harmonic oscillator $O$ via a Jaynes-Cummings (JC) 
Hamiltonian, an assumption frequently performed in cavity quantum electrodynamics (QED) (see, for instance, \cite{meystre,lofrancoCQED}). 
Qubit $B$ is virtually isolated from any environment.
Initially, the two qubits are prepared in the Bell state $\ket{\phi_{AB}^+}$
and $O$ in its ground state $\ket{0_O}$, see Fig.~\ref{fig:JC-interaction}(a) left. 
In the interaction picture, 
the Hamiltonian is ${\cal H}_{AO}=g(\sigma_+ a + \sigma_- a^\dag)$,
where $g$ is the coupling constant, $\sigma_+$ ($\sigma_-$) the qubit raising (lowering) operator, 
and $a^\dag$ ($a$) the oscillator creation (annihilation) operator. 
At \textbf{time} $\mathbf{\bar{t}}=\pi/g$ the states of $A$ and $O$ are swapped  
with respect the initial state, and the global state becomes
$\ket{0_A}\otimes\ket{\phi_{BO}^+}$, see Fig.~\ref{fig:JC-interaction}(a) right.
We have that $\rho_{AB}(\bar{t})=\ket{0_A}\otimes\frac{1}{2}\openone_B$,
the $A$-$B$ entanglement is zero, being completely 
transferred to $B$-$O$.
At time $\bar{t}$, any unravelling of the $A$-$B$ dynamics (see 
\ref{quantum-ensemble-time-evolutions-qenv}) gives a quantum 
ensemble whose average entanglement is zero, so that $E_h(\bar{t})=0$:
at this time no classical communication or local operation can help to recover any 
entanglement between $A$ and $B$. Only the subsequent interaction between $A$ and $O$ 
can gradually restore the $A$-$B$ entanglement: At \textbf{time} $\mathbf{2\bar{t}}$,
when a new $A$-$O$ swapping is completed, the initial state is just retrieved.
Therefore, the entanglement revival is here due to the 
perfect {\em entanglement back-transfer},
as well-known in the literature~\cite{bellomo2007PRL,lopez2008PRL}.

This explanation is unsuitable for the examples considered above in this paper,
where the environment is classical and no entanglement transfer is possible. 
In those cases, 
during the dynamics, the environment acquires only classical information about the system $A$-$B$ and
does not entangle with $A$ or $B$. Quantum correlations do not leave the system $A$-$B$ but 
they are simply not accessible due to the lack of classical information.
Indeed, if at time $\bar{t}$ when $E_h(\bar{t})=1$, someone provides $A$-$B$ with the 
classical information about which random unitary the system 
underwent (in the case of random local fields) or about which random 
phase is added to the system states (in the case of stochastic pure-dephasing noise), 
then all the $A$-$B$ entanglement can be recovered. 

In order to get further insight on the phenomenon occurring in the quantum system $ABO$,
here we estimate the hidden entanglement between $A$ and $B$. 
To this end we suppose to perform a measurement of the system $O$  
in the orthonormal basis $\{\ket{0},\ket{1}\}$\footnote{The physical 
decomposition of the $AB$ system interacting with another quantum system in 
general is not unique, see \ref{quantum-ensemble-time-evolutions-qenv}. The
decomposition we choose may be physically realizable in cavity QED
systems, moreover one can numerically check that it gives the largest amount of
average entanglement~\cite{mascarenhas11}.}. 
Under these conditions the physical quantum ensemble describing the $AB$ quantum dynamics reads 
${\cal A}=\{\big(p_0(t), \ket{\varphi_0(t)}\big),\big(p_1(t), \ket{\varphi_1(t)}\big)\}$, where
\begin{equation}
 p_0(t)= \frac{1}{2}\big(1+\cos^2(gt/2)\big),
 \quad p_1(t)=\frac{1}{2}\sin^2(gt/2)
 \label{eq:JC-QE-probabilities}
\end{equation}
are respectively the probability that the system $O$ is found in $\ket{0}$ or in $\ket{1}$,
and
\begin{equation}
 \ket{\varphi_0(t)}= \frac{1}{\sqrt{2p_0(t)}}\big(\ket{00}+\cos(gt/2)\ket{11}\big),
  \quad \ket{\varphi_1(t)}=\ket{01}
 \label{eq:JC-QE-states}
\end{equation}
are the corresponding system $AB$ states. Note that only the state 
$\ket{\varphi_0(t)}$ is an entangled state.
Therefore the average entanglement of the quantum ensemble $\cal A$ is given by:
\begin{equation}
 E_{av}({\cal A},t)= p_0(t) E(\ketbra{\varphi_0(t)}{\varphi_0(t)})\,=
  \frac{1+\eta}{2}\mathtt{f}\Big(\frac{2\sqrt{\eta}}{{1+\eta}}\Big)
\label{eq:JC-average-entanglement}
\end{equation}
and the hidden entanglement reads
\begin{equation}
 E_{h}({\cal A},t)=E_f(\rho_{AB}(t))\,-\,E_{av}({\cal A},t)=\, \mathtt{f}\big(\sqrt{\eta}\big)\,-\, 
  \frac{1+\eta}{2}\mathtt{f}\Big(\frac{2\sqrt{\eta}}{1+\eta}\Big),
\label{eq:JC-hidden-entanglement}
\end{equation}
where we set $\mathtt{f}(x) \equiv \mathtt{h}(\frac{1+\sqrt{1-x^2}}{2})$, with the function $\mathtt{h}$ 
defined below Eq.~(\ref{eq:wootters-formula}), and $\eta=\cos^2(gt/2)$.
In Fig.~\ref{fig:JC-interaction}(b) we show $E_{av}({\cal A},t)$ (thin blue line), 
the entanglement of formation  $E_f(\rho_{AB}(t))$  (tick red line)
and the corresponding hidden entanglement 
$E_h({\cal A}(t))$ (dashed line)\footnote{It is worth to notice that
Eq.~(\ref{eq:JC-hidden-entanglement}) also gives the hidden entanglement associable
to an \textit{amplitude damping 
channel}~\cite{nielsen-chuang,benenti-casati-strini} applied to the qubit $A$, where $1-\eta$
is the probability that $A$ looses a photon.}.
We observe the existence of a small amount of HE
at times $gt \neq\pi$. 
It represents the  extra (with respect to the entanglement of formation) amount of
entanglement that it would be possible to recover if the classical information
coming from measurements of the quantum state of $O$ would be available. 
The fact that $E_h({\cal A}(t))$  is much smaller than the initially present entanglement
indicates that the main mechanism underlying the decrease (recover) of $AB$ entanglement, it is not the loss 
(gain) of classical information on the system $AB$, but it is rather the development 
(regression) of quantum correlations between $B$ and $O$.

A few remarks are now in order. Similar entanglement revivals can be observed when the quantum 
harmonic oscillator interacts with an environment inducing a non-Markovian dynamics of $A$-$B$ 
($O$ plus its environment representing a structured bath acting locally on $A$ and influencing
nontrivially the quantum evolution of $AB$). In this case, the entanglement recovery signals
the non-Markovian quantum evolution of one of the parts of the bipartite system $AB$ acted by
a {\em quantum environment}, as discussed in Ref. \cite{rivas}.
The example of the two qubits affected by low-frequency noise 
highlights that when a bipartite quantum system is affected by  
{\em classical non-Markovian noise}, the induced non-Markovian dynamics of the bipartite system 
does not justify by itself the occurrence of entanglement 
revivals between the independent subsystems. Indeed, in 
Fig.~\ref{fig:PureDeph-echo} the dynamics without $\pi$-pulses is
non-Markovian and entanglement revivals do not appear. 
We may summarize the above observations saying that the origin of entanglement revivals displayed 
by a noninteracting bipartite quantum system may be either due
to (i) a back-and-forth transfer of quantum correlations with a quantum environment possibly acting locally on one
of the two subsystems and inducing a non-Markovian dynamics, or to (ii) 
the action of local operations on one subsystem affected by a classical 
non-Markovian noise source. In this last
case the recovery is just the manifestation of quantum correlations which remain ``hidden'' in
the quantum system, i.e. not directly available in the density matrix description of the system state.

\section{Discussion}
\label{sec:discussion}
From the examples above we argue that recovery of entanglement is achievable 
without nonlocal operations,
when various members of the physical ensemble evolve differently from each other but unitarily. 
In this case, even though the evolution of the ensemble averaged density matrix is not unitary, 
no qubit-environment entanglement is generated.
Using a terminology borrowed from NMR~\cite{cory} 
we speak of \emph{incoherent errors}, whereas 
\emph{decoherent errors} arise when
the evolution is non-unitary even for a single member of 
the ensemble. For this latter case the average entanglement decays. 

We point out that HE depends on the quantum ensemble \textit{physically} 
giving the system state. The key point is that the physical dynamics subsumes a 
specific decomposition for the evolved density matrix of the system.
In the above examples relative to random local fields and classical non-Markovian noise,
such physical decomposition is always an ensemble 
of maximally entangled states.
Had we considered a dynamics such as to give, at $t=\bar{t}$, a mixture of separable states 
as decomposition of the system density matrix, 
no local operation would have been capable to recover entanglement ($E_h(\bar{t})=0$).
The dynamics of the system after the application of the pulse proves 
that, at time $\bar{t}$, the two decompositions are not equivalent. 

Finally, we remark that the results of the previous examples do not
violate the monotonicity 
axiom~\cite{PlenioReview,HorodeckiReview,Bennett96}:
\textit{entanglement cannot increase under LOCC}.
The key point is that this axiom is fulfilled by all
entanglement measures provided we consider local operations
which are completely positive, trace preserving (CPT) maps. 
On the other hand, not all physical operations result in a composition 
of completely positive trace preserving maps across successive time intervals: 
there exist non-divisible dynamical maps, as discussed in Refs.~\cite{breuer,rivas}.
In the previous examples entanglement recovery, from 
time $\bar{t}$ to time 2$\bar{t}$, is induced by purely
local operations which are not LOCC, 
since the corresponding density matrix
evolution cannot be described by a CPT map. 
To prove this point, it is enough to observe that 
the density matrix is such that $\rho(2\bar{t})=\rho(0)$. Therefore
driving the system from the state $\rho(\bar{t})$ to 
$\rho(2\bar{t})$ is equivalent to driving it to 
$\rho(0)$. This operation cannot be described by a CPT
map since the (CPT) evolution from time $0$ to time $\bar{t}$
is not invertible.

\section{Conclusions}
\label{sec:conclusions}
In conclusion, we have introduced the concept of hidden entanglement (HE) based on 
the description of the system dynamics in terms of the ensemble of pure states 
physically underlying the system time evolution.
We used the concept of HE to give a physical explanation on the phenomenon of 
entanglement revivals, in those cases in which the system is subject to 
classical noise sources.
We showed as there is no violation of the entanglement monotonicity axiom 
because no entanglement is destroyed and created in this case: a nonzero HE 
signals a loss of entanglement that is not due to the establishment of 
quantum correlations with the environment.
Quantum correlations remain within the system, but they are not exploitable 
due to the lack of classical information.

The concept of HE can be also applied when a system interacts with a quantum environment.
In these cases, however, evaluation of HE requires some 
environment ``monitoring strategy''~\cite{Nhal04,Carvalho07,mascarenhas11,gregoratti,GMDA,volgelsberger,barchielli,murch},
in order to realize a statistical 
ensemble of pure state evolutions which \textit{physically} underlies the system dynamics.   

We stress that our analysis has direct application to solid state nanodevices
which prevalently suffer from low-frequency noise. 
Indeed HE 
may be a figure of merit
indicating the amount of entanglement resources which can be recovered
by using local sequences of standard pulses, an appealing feature for
quantum control in distributed architectures 
for quantum computing where different 
subunits are subject to different non-Markovian noise sources.
This allows us
to avoid resorting to non-local control~\cite{mukhtar2010PRA1}, 
which may be a much more demanding task. 

\ack
A.D. thanks Laura Mazzola for useful discussions. 
R.L.F. thanks Giuseppe Compagno for fruitful discussions. 
This work was partially supported by 
the European Community through grant no. ITN-2008-234970 NANOCTM
and by PON02-00355-339123 - ENERGETIC.
G.B. acknowledges support by
MIUR-PRIN project \emph{Collective quantum phenomena:
From strongly correlated systems to quantum simulators}.

\appendix

\section{Quantum ensembles and system dynamics}

\subsection{Evolution in the presence of classical noise}
\label{quantum-ensemble-time-evolutions}

In this Appendix we demonstrate
that when a system is affected by \textit{classical noise sources}
its dynamics admits a single physical decomposition in terms  of an ensemble of pure state evolutions. 
To this end let us consider a quantum system $Q$ that evolves according to the following Hamiltonian:
\begin{equation}
{\cal H}_{Q}(t)= {\cal H}_0 \,+\,\hat{q} \, x(t),
\label{appendix:semiclassical-hamiltonian}
\end{equation}
where ${\cal H}$ and $\hat{q}$ are Hermitian operators acting on the system's Hilbert space, and
$x(t)$ is a stochastic process representing the effect of a classical noise source.
The system quantum evolution 
for a given realization of ${x}(t)$ is expressed in terms of the evolution operator
\begin{equation}
U_{Q}\big(t|x(t)\big)=\hat{T}e^{-i\int_0^t dt'{\cal H}_{Q}(t')} \,.
\label{appendix:UQ-operator}
\end{equation}
Assuming that the system is initially prepared in a pure state 
$\ket{\psi(t=0)}$ $=\ket{\psi_0}$,
the system state at a generic time $t$ is 
\begin{equation} 
\big|\psi\big(t|x(t)\big)\big\rangle=U_Q\big(t|{x}(t)\big)\ket{\psi_0} \, .
\end{equation}
The system evolution is represented by the quantum ensemble 
\begin{equation}
{\cal A}=\Big\{P[x(t)],\,\big|\psi\big(t|x(t)\big)\big\rangle\Big\}
\label{appendix:quantum-ensemble-1}
\end{equation}
where $P[x(t)]$ is the probability of a given realization $x(t)$.
The system density matrix is obtained by averaging over the quantum ensemble
(\ref{appendix:quantum-ensemble-1}), and it is expressed as a path integral:
\begin{equation}
\rho(t)=\int D[x(t)] P[x(t)]\,\big|\psi\big(t|x(t)\big)\big\rangle
                           \big\langle\psi\big(t|x(t)\big)\big|\,.
\label{appendix:density-operator-1}
\end{equation}
This description applies to several scenarios in the solid state~\cite{Slichter,PaladinoRMP2013}.
In relevant situations the stochastic process $x(t)$ can be consider static during
the time evolution (see for instance \cite{FalciPrl05}). 
In these cases $x(t)$ can be replaced by a random 
variable $x$ and the probability $P[x(t)]$ is replaced with
the probability density function $p(x)$. The path-integral 
(\ref{appendix:density-operator-1}) reduces to an ordinary integral:
\begin{equation}
\rho(t)=\int dx\, p(x)\, \ketbra{\psi_x(t)}{\psi_x(t)}, \qquad {\cal A}=\{p(x)dx,\ket{\psi_x(t)}\},
\label{appendix:quantum-ensemble-2}
\end{equation}
where $\ket{\psi_x(t)}=\ket{\psi(t|x)}$.
If  $x$ is a discrete random variable the quantum ensemble reduces to the general form 
introduced in Section~\ref{sec:HE}
\begin{equation}
{\cal A}=\{p_i,\ket{\psi_i(t)}\}, \qquad \rho(t)=\sum_i p_i \ketbra{\psi_i(t)}{\psi_i(t)},
\label{appendix:quantum-ensemble-3}
\end{equation}
where $p_i \equiv p(x=x_i)$, and 
$\ket{\psi_i(t)}=\ket{\psi(t|x_i)}$.

\subsection{Evolution in the presence of a quantum environment}
\label{quantum-ensemble-time-evolutions-qenv}

The above description in terms of quantum ensemble is no longer unique when 
the quantum system interacts with
a quantum environment.  To illustrate this fact let us suppose that
the quantum system is coupled to a quantum environment $E$ with a total Hamiltonian
for $Q$ and $E$ given by
\begin{equation}
{\cal H}_{QE}= {\cal H}_{0}\,+\,\hat{q}\otimes \hat{X}_E+{\cal H}_E,
\label{appendix:QE-Hamiltonian}
\end{equation}
where ${\cal H}_{0}$ and $\hat{q}$ are the same system $Q$ operators which appear in 
(\ref{appendix:semiclassical-hamiltonian}), $\hat{X}_E$ is an Hermitian
operator for the environment $E$ and ${\cal H}_E$ is free evolution Hamiltonian for the environment.
The system-plus-environment  evolves unitarily with
\begin{equation}
U_{QE}(t)=e^{-i{\cal H}_{QE}t}.
\label{appendix:UQE-operator}
\end{equation}
For the sake of simplicity here we assume that the system $QE$ is initially in the state
$\ket{\Psi_{QE}(0)}=\ket{\psi_0} \otimes \ket{\xi_E}$~\cite{nielsen-chuang,benenti-casati-strini}.
The system $QE$ state at time $t$ is:
\begin{equation}
\ket{\Psi_{QE}(t)}=U_{QE}(t)\ket{\psi_0}\otimes\ket{\xi_E}.
\label{appendix:QE-state}
\end{equation}
The reduced density matrix of $Q$ is
\begin{equation}
\rho_{Q}\,=\,\textrm{Tr}_E\Big\{U_{QE}(t)\,\ketbra{\psi_0}{\psi_0}\otimes\ketbra{\xi_E}{\xi_E}\, U^\dag_{QE}(t)\Big\}.
\label{appendix:Q-density-matrix-1}
\end{equation}
The trace operation on the environment $E$ can be carried out by choosing
an orthonormal basis $\kappa=\{\ket{k_E}\}$ for $E$:
\begin{eqnarray}
&&\rho_{Q}\,=\,\sum_{ k } \bra{k_E} \Big(U_{QE}(t)\,\ketbra{\psi_0}{\psi_0}\otimes
               \ketbra{\xi_E}{\xi_E}\, U^\dag_{QE}(t)\Big)\ket{k_E}\nonumber\\
&&\hspace{0.6cm}=\,\sum_{k } \bra{k_E} U_{QE}(t) \ket{\xi_E} \, \ketbra{\psi_0}{\psi_0}
               \bra{\xi_E}\, U^\dag_{QE}(t)\ket{k_E}\nonumber\\
&&\hspace{0.6cm}=\,\sum_{k } U_{Q_k}(t)  \, \ketbra{\psi_0}{\psi_0}
               \, U^\dag_{Q_k}(t),\nonumber\\
\label{appendix:Q-density-matrix-2}
\end{eqnarray}
where $U_{Q_k}(t)\equiv \bra{k_E} U_{QE}(t) \ket{\xi_E}$ are system $Q$ operators.
Note that in general $U_{Q_k}(t)$ are not unitary operators. By defining
the normalized $Q$ state 
\begin{equation}
  \ket{\psi_k(t)}=\frac{1}{\sqrt{\eta_k}} U_{Q_k}(t) \ket{\psi_0},
\label{appendix:Q-conditioned-pure-state}
\end{equation}
where
\begin{equation}
{\eta_k}=\bra{\psi_0}U^\dag_{Q_k}(t)U_{Q_k}(t) \ket{\psi_0},
\label{appendix:Q-conditioned-pure-state-probability}
\end{equation}
the reduced density matrix for $Q$ reads
\begin{equation}
  \rho_Q(t)=\sum_k \eta_k \ketbra{\psi_k(t)}{\psi_k(t)}.
\label{appendix:Q-density-matrix-3}
\end{equation}
The trace operation corresponds to perform a measurement of the system $E$ with respect 
to the orthonormal basis $\kappa=\{\ket{k_E}\}$. 
When the outcome of the measurement is $k$ then the 
system $Q$ is in the pure state $\ket{\psi_k(t)}$.
Therefore, 
the system $Q$ is described by the quantum ensemble
\begin{equation}
{\cal A}_\kappa =\{\eta_k,\ket{\psi_k(t)}\},
\label{appendix:quantum-ensemble-conditioned}
\end{equation}
where $\eta_k$ is the probability to obtain the outcome $k$, given by 
(\ref{appendix:Q-conditioned-pure-state-probability}).
We have unravelled the $Q$ dynamics into a statistical ensemble 
of pure state evolutions, the so-called 
quantum trajectories~\cite{carmichael}.

The measurement of the environment $E$ removes the arbitrariness of the $\rho_Q$ decomposition,
the system $Q$ being \textit{physicraally described}~\cite{Nhal04,Carvalho07,
mascarenhas11,volgelsberger,mascarenhas10} by 
the ensemble (\ref{appendix:quantum-ensemble-conditioned}). Once the classical information
about the   outcomes $k$ is known, the average entanglement $E_{av}({\cal A}_{{k}})$ can be 
obtained by means of local operations.
Also in this case, hidden entanglement represents the amount of entanglement
which can be recovered, once the classical information is provided to the system.
We remark that in general the above information about the environment is
not easily accessible,
since the environment is supposed to represent an ensemble of
a huge number of uncontrollable degrees of freedom.
Nevertheless, under limiting situations it may be possible
to have some control on a selection of those degrees of freedom~\cite{murch}.  

The above considerations also point out that when a system interacts with a quantum environment, 
the average entanglement (and consequently the hidden entanglement) depends on the adopted environment 
monitoring strategy. 
One can in principle look for the system ensemble with the largest amount of average 
entanglement; this maximum entanglement is called \textit{entanglement of assistance}~\cite{divincenzo} or 
\textit{localizable entanglement}~\cite{verstraete}.
This ``best'' ensemble is not always achievable, since one has to deal with physically 
realizable measurements~\cite{Carvalho07,mascarenhas11}.

\end{document}